\documentstyle[preprint,aps]{revtex}

\begin{document}
\def \cosw{\cos \chi_w^t}
\def \sinw{\sin \chi_w^t}
\def \ce{\cos \chi_e^w}
\def \se{\sin \chi_e^w}
\def \sse{\sin^2 \chi_e^w}
\def \pew{\Phi}
\def \chie{\chi_e^w}
\def \chiw{\chi_w^t}
\def \sw{\sin \theta_W}
\def \ssqw{\sin^2 \theta_W}
\def \cw{\cos \theta_W}
\def \thetad{\theta^{\dagger}}
\def \thetas{\theta^*}
\def \sangle{\xi}
\def \ds{\frac{d \sigma}{d \cos \theta^*}}
\def \norm{ \Bigl( ~\frac{3 \pi \alpha^2}{2 s}\beta ~\Bigr) ~}
\def \norme{\Bigl( ~\frac{3 \pi \alpha^2}{8 s}\beta ~\Bigr) ~}
\def \norms{\Bigl( ~\frac{ \pi \alpha_s^2}{9 s}\beta ~\Bigr) ~}
\def \beq{\begin{equation}}
\def \eeq{\end{equation}}
\def \beqa{\begin{eqnarray}}
\def \eeqa{\end{eqnarray}}
\def \amp{{\cal M}}
\def \ttbar{$t \bar t$}
\def \qqbar{$q \bar q$}
\def \antibeamline{``anti-beamline''}
\def \beamline{Beamline}
\def \UP{{\rm R}}
\def \DOWN{{\rm L}}
\def \up{\uparrow}
\def \down{\downarrow}
\def \half{\hbox{$1\over2$}}
\def \qbar{\bar q}
\def \tbar{\bar t}
\def \ebar{\bar e}
\def \opt{Off-diagonal}
\def \epm{$e^+e^-$}
\def \eL{$e^-_L~e^+_R$}
\def \eR{$e^-_R~e^+_L$}
\def \UD{$t_{\up} ~{\bar t}_{\down}$ }
\def \DU{$t_{\down} ~{\bar t}_{\up}$ }
\def \UU{$t_{\up} ~{\bar t}_{\up}$ }
\def \DD{$t_{\down} ~{\bar t}_{\down}$ }


\def	\nn		{\nonumber}
\def	\=		{\;=\;} 
\def	\ret		{\\[\eqskip]}
\def	\to		{\rightarrow }

\draft
\preprint{
  \parbox{2in}{Fermilab--Pub--96/042-T \\
  [-0.12in] hep-ph/9606419
}  }

\title{Spin Correlations in Top-Quark Pair Production \\at $e^+e^-$ Colliders}
\author{Stephen Parke \cite{SPemail} and Yael Shadmi \cite{YSemail}}
\address{Theoretical Physics Department\\
Fermi National Accelerator Laboratory \\
P.O. Box 500, Batavia, IL  60510 }
\date{June 21, 1996}   
\maketitle
\begin{abstract}
We show that top-quark pairs are 
produced in an essentially unique spin configuration
in polarized $e^+e^-$ colliders at all energies
above the threshold region. 
Since the directions of the electroweak  decay products 
of polarized top-quarks are strongly
correlated to the top-quark spin axis, this unique spin configuration
leads to a  distinctive topology for top-quark pair events 
which can be used to 
constrain anomalous couplings to the top-quark.
A significant interference effect between the {\it longitudinal} and 
{\it transverse} W-bosons in the decay of polarized top-quarks
is also discussed.
These results are obtained at leading order in perturbation theory but
radiative corrections are expected to be small.
\end{abstract}

\newpage

Since the top-quark, with mass in the range of 
175 GeV~\cite{CDFtop,D0top}, 
decays electroweakly
before hadronizing~\cite{Bigi},
there are significant angular correlations between
the decay products of the top-quark and the spin of the top-quark.
Therefore,
if top-quark pairs are produced with large polarizations,
there will be sizable
angular correlations between the decay products 
and the spin orientation of each top-quark, 
as well as between the decay products of the top-quark
and the decay products of the top anti-quark.
These angular correlations depend sensitively on the top-quark couplings
to the $Z$ and the photon, and to the $W$ and $b$-quark. 
Many authors have proposed to use the angular information in top-quark 
events produced at \epm\ colliders to constrain deviations 
from Standard-Model 
couplings~[\ref{nlcstart}-\ref{nlcend}]. 
In most of these studies, the top-quark spin is decomposed   
in the helicity basis, i.e. along the direction of motion of the quark.
Recently, Mahlon and Parke~\cite{MP2} have shown that this decomposition
in the helicity basis is far from the optimal decomposition 
for top-quarks produced at the Tevatron.
In this paper we address the issue of what is the optimal
decomposition of the top-quark spins for  \epm\ colliders.
Here, we only consider top-quark production at energies that are above
the threshold region~\cite{threshold}.
We give the differential cross-section for top-quark pair 
production for a generic spin basis.
We have found that there is a decomposition for which the 
top-quark pairs are produced in an essentially unique spin configuration at
polarized  \epm\ colliders. We call this spin basis the 
``\opt'' basis because the contribution from like-spin pairs
vanishes to leading order in perturbation theory.
Finally, we discuss the angular correlations between the decay products
of the top-quarks, and point out a significant interference effect
between the {\it longitudinal} and 
{\it transverse} W-bosons in the decays of the polarized top-quarks.

\begin{figure}[h]
\vspace*{20cm}
\includegraphics{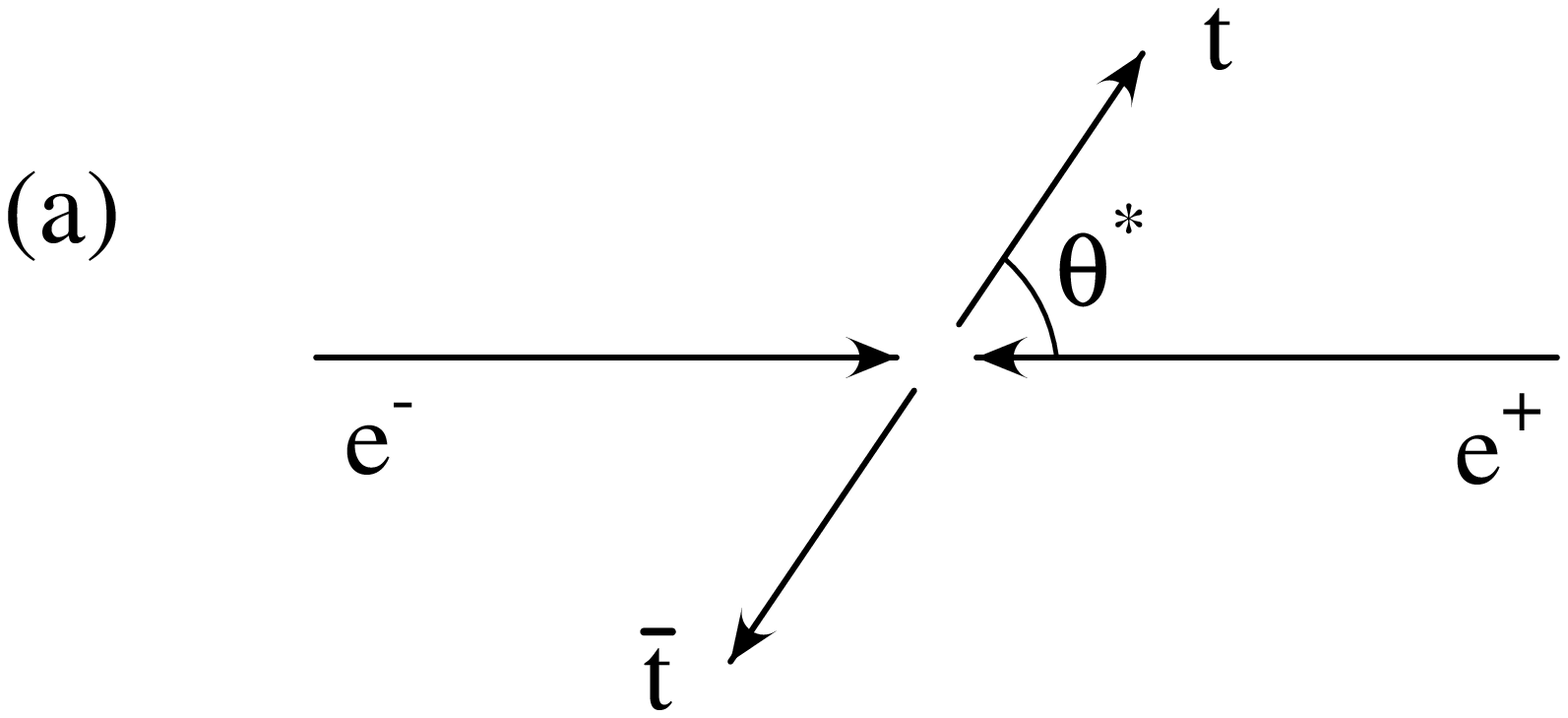}
\includegraphics{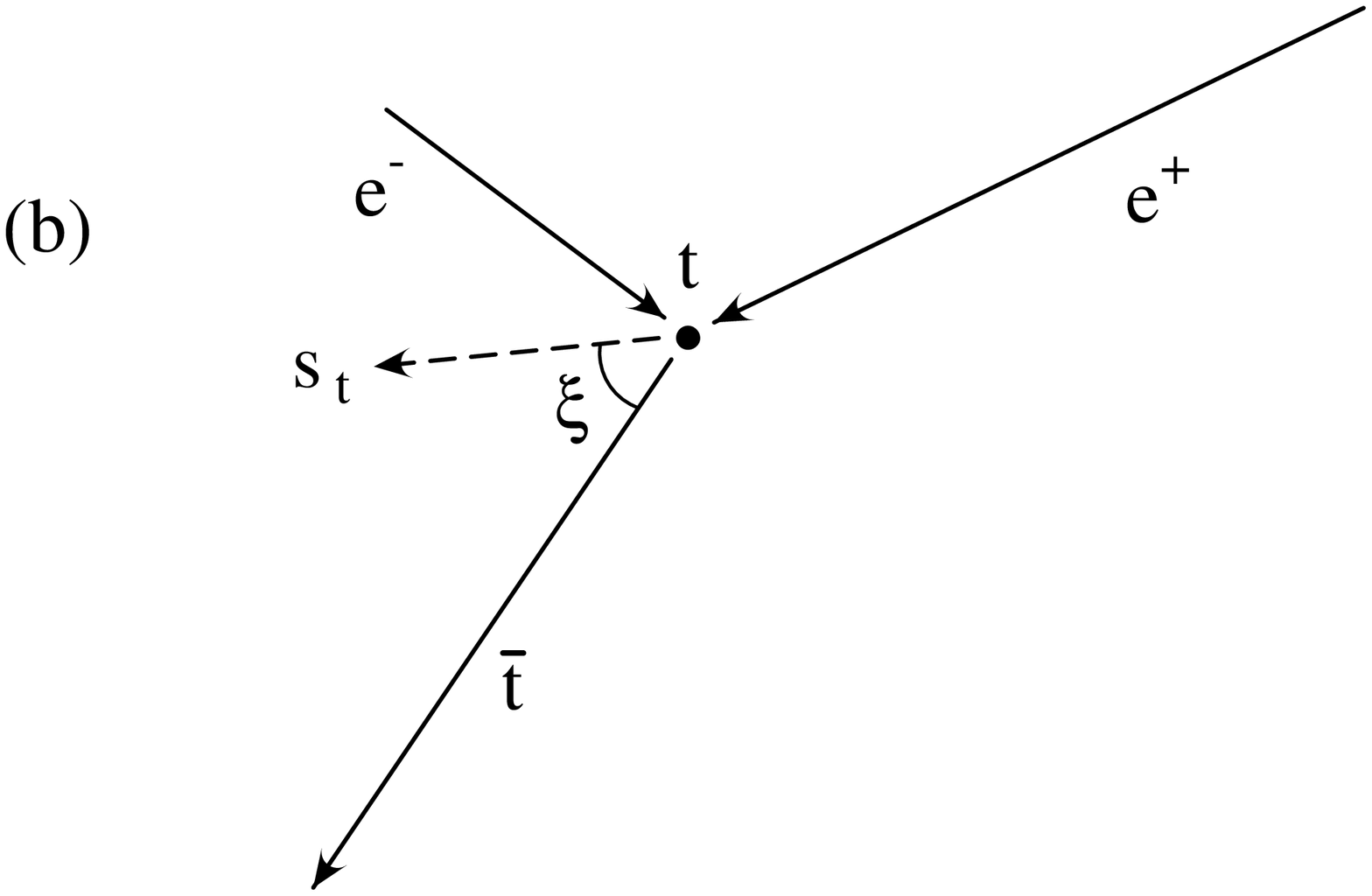}
\includegraphics{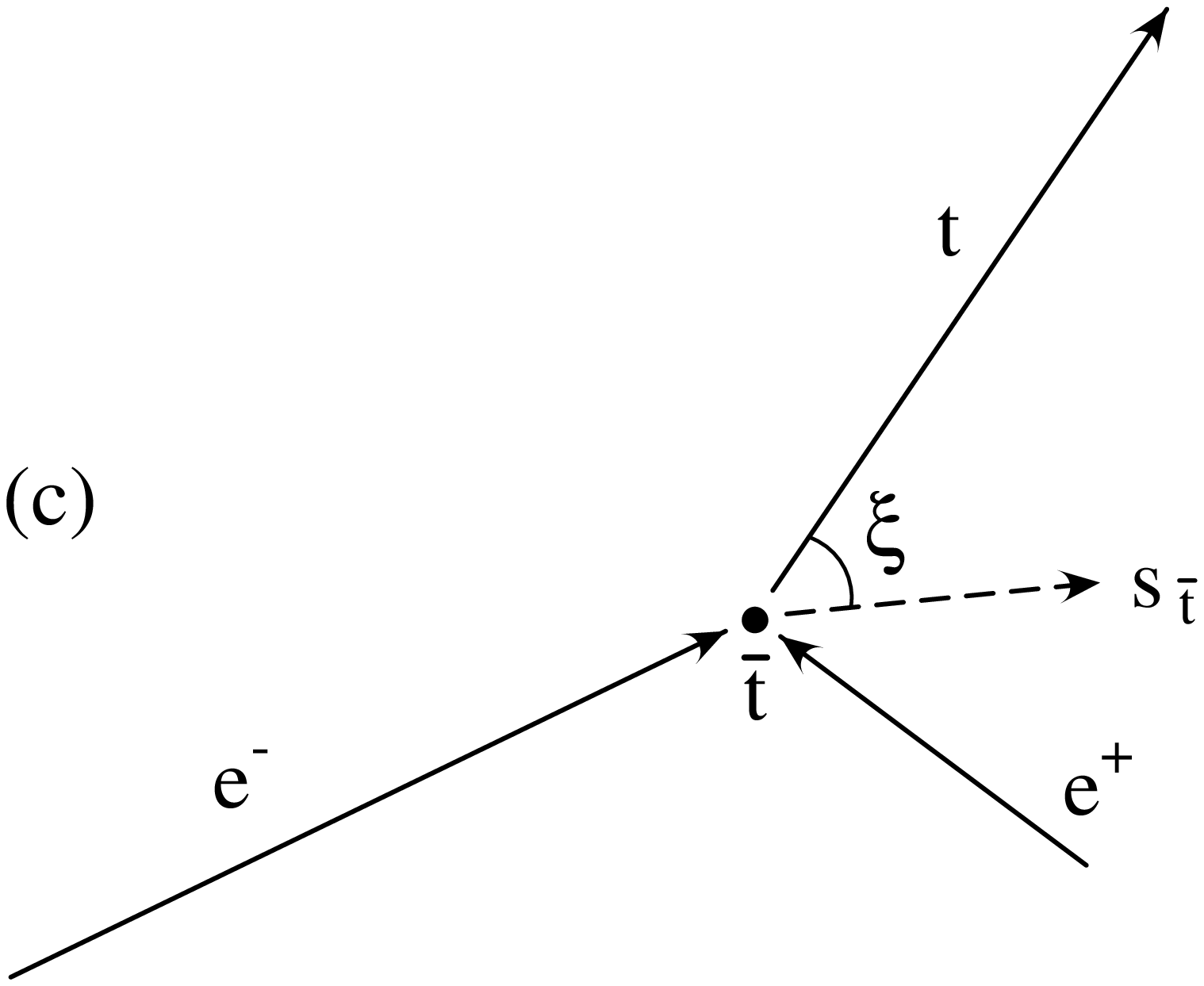}
\vspace{1.0cm}

\caption[]{
The scattering process in the center-of-mass frame (a), 
in the top-quark rest-frame (b) and in the top anti-quark rest-frame (c). 
${\bf s_t}$ (${\bf s_{\bar t}}$) 
is the top (anti-top)  spin axis. 
}
\label{frames}
\end{figure}

The generic spin basis we will consider in this paper 
is shown in Figure~\ref{frames}. 
This choice is motivated by two features of the Standard Model prediction for 
top-quark pair production at  \epm\ colliders at leading-order in 
perturbation theory;
the transverse top-quark polarization is zero, 
and CP is conserved.
Therefore, we have defined our general basis so that 
the spins of the top quark and anti-quark are in the production plane,
so there is no transverse polarization,
and the spin four-vectors are back-to-back in the zero momentum frame, 
so that states with opposite spins are CP eigenstates.

The top-quark spin states are defined in the top-quark rest-frame, 
where we decompose the top spin along the direction ${\hat s}_t$, 
which makes an angle $\sangle$ with the 
anti-top momentum in the clockwise direction.
Similarly, the top anti-quark spin states are defined in the anti-top
 rest-frame, 
along the direction ${\hat s}_{\bar t}$, 
which makes the {\it same} angle $\sangle$ with the top momentum
also in the clockwise direction.
Thus, the state 
$t_{ {\up}}  {\bar t}_{ {\up}}$ 
($t_{ {\down}}  {\bar t}_{ {\down}}$ )
refers to a top with spin in the $+{\hat s}_t$ ($-{\hat s}_t$) 
direction in the top rest-frame, 
and an anti-top with spin  
$+{\hat s}_{\bar t}$ ($-{\hat s}_{\bar t}$) in the anti-top rest-frame.

Using this generic spin basis,
the Standard Model leading-order polarized cross-sections 
for top-quark pair production 
at center-of-mass energy $\sqrt{s}$, top-quark speed $\beta$ 
and top-quark scattering angle $\thetas$, are given by
\beqa
\ds ~(e^-_L~e^+_R \rightarrow t_{\up}~\tbar_{\up}) & = &
\ds ~(e^-_L~e^+_R \rightarrow t_{\down}~\tbar_{\down})
\nn \\ &  = & 
\norm |A_{LR} \cos \sangle ~-~ B_{LR} \sin \sangle ~|^2 \ , \nn \\[0.2in]
\ds ~(e^-_L~e^+_R \rightarrow t_{\up}~\tbar_{\down} 
~or~ t_{\down}~\tbar_{\up}) & = &
\norm |A_{LR} \sin \sangle ~+~ B_{LR} \cos \sangle ~\pm~ D_{LR}|^2.
\label{LRxsec} \,
\eeqa
Here $\alpha$ is the QED fine structure constant 
and the quantities A, B and D
depend on the kinematic variables $\beta$ and $\cos \thetas$,
and  on the 
fermion couplings to the photon and Z-boson, in the following way
\beqa
A_{LR} & = & [(f_{LL}+f_{LR}) ~\sqrt{1-\beta^2} ~\sin \thetas]/2 \ , \nn \\
B_{LR} & = & [f_{LL}~(\cos \thetas 
+ \beta)+f_{LR}~(\cos \thetas - \beta)]/2 \nn  \ ,  \\
D_{LR} & = & [f_{LL}~(1 + \beta \cos \thetas) 
+ f_{LR}~(1 - \beta \cos \thetas)]/2   \ ,
\label{ABDdefn}
\eeqa
with
\beqa
f_{IJ} & = & Q_{\gamma}(e) Q_{\gamma}(t) 
+ Q_{Z}^I(e) Q_{Z}^J(t)\Bigl( \frac{1}{\ssqw} \Bigr)
\Bigl( \frac{s}{(s-M_Z^2)+iM_Z\Gamma_Z}\Bigr) \ ,
\label{fdefn}
\eeqa
where  $M_Z$ is the $Z$ mass, $\Gamma_Z$ is the $Z$ width,
and $I,J ~ \in ~(L,R)$. 
The electron couplings are given by
\beq
	Q_{\gamma}(e)=-1, 
\quad    	Q_Z^L(e)=\frac{2 \ssqw -1}{2\cw},
\quad 	Q_Z^R(e)= \frac{\ssqw}{\cw} \ , 
\eeq
and the top-quark couplings are given by
\beq
	Q_{\gamma}(t)=\frac{2}{3},
\quad    	Q_Z^L(t)=\frac{3-4\ssqw}{6\cw},  
\quad   	Q_Z^R(t)=\frac{-2\ssqw}{3\cw} \ .
\eeq
The angle $\theta_W$ is the Weinberg angle.
In the limit $s \gg M_Z^2$ 
the products of fermion couplings
for top-quark production 
are
\beq
f_{LL}=-1.19, \quad f_{LR}=-0.434, 
\quad f_{RR}=-0.868, \quad f_{RL}=-0.217 \  ,
\eeq
where, as  throughout this paper, we use $\ssqw = 0.232$.

The cross-sections for $e^-_R~e^+_L$ may be obtained 
from eqns.~(\ref{LRxsec}-\ref{fdefn})
by interchanging
$L$ and $R$ as well as $\up$ and $\down$.
These cross-sections can be conveniently 
derived using the spinor helicity method for massive 
fermions described in~\cite{MP2}.

Our generic basis reproduces the familiar helicity basis
for the special case 
\beq
\cos \sangle = \pm 1 \ ,
\label{helangle}
\eeq
for which the top-quark spin is defined along its 
direction of motion.
Substituting eqn.~(\ref{helangle}) in our general polarized cross-section
expressions~(\ref{LRxsec}), we recover
the well known helicity cross-sections
\beqa
\ds ~(e^-_L~e^+_R \rightarrow t_L~\tbar_L) & = &
\ds ~(e^-_L~e^+_R \rightarrow t_R~\tbar_R) \nn \\
& = & \norme |f_{LL}+f_{LR}|^2 ~(1-\beta^2) \sin^2 \thetas \ , \nn \\[0.1in]
\ds ~(e^-_L~e^+_R \rightarrow t_R~\tbar_L
~or~ t_L~\tbar_R) & = &
\norme |f_{LL}(1 \mp \beta) + f_{LR}(1 \pm \beta)|^2 ~(1\mp \cos \thetas)^2 \ .
\label{Helicity}
\eeqa

Another useful basis is the ``Beamline basis''~\cite{MP2}, 
in which the top-quark spin axis is the 
{\it positron direction in the top rest-frame},
and the top anti-quark spin axis is the electron direction in the 
anti-top rest-frame.
In terms of the spin angle $\sangle$, this corresponds to   
\beq
\label{beamlinesubs}
\cos \sangle = \frac{ \cos \thetas + \beta}{1 + \beta \cos \thetas}\ . 
\eeq
\newpage
The polarized cross-sections in this basis are 
\beqa
 \ds ~(e^-_L~e^+_R \rightarrow t_{\up}~\tbar_{\up})  & = &
\ds ~(e^-_L~e^+_R \rightarrow t_{\down}~\tbar_{\down}) \nn \\
& = & \norm
|f_{LR}|^2 ~\frac{\beta^2(1-\beta^2) \sin^2 \thetas }
{(1+\beta \cos \thetas)^2} 
\ , \nn \\[0.1in]
\ds ~(e^-_L~e^+_R \rightarrow t_{\down}~\tbar_{\up} ) & = & 
\norm |f_{LR}|^2~\frac{\beta^4 \sin^4 \thetas}{(1+\beta \cos \thetas)^2} 
\ , \nn \\[0.1in]
\ds ~(e^-_L~e^+_R \rightarrow t_{\up}~\tbar_{\down} ) & = &
\norm \Biggl| f_{LL}(1+\beta \cos \thetas)
+f_{LR}\frac{(1-\beta^2)}{(1+\beta\cos\thetas)} \Biggr|^2 \ .
\label{Beamline}
\eeqa
Note that in this basis,  three out of the four polarized cross-sections  
are proportional
to $|f_{LR}|^2$ which is much smaller $|f_{LL}|^2$.
These three components are further 
 suppressed by at least two powers of $\beta$.
The remaining component, $t_{\up}\bar t_{\down}$, will
therefore account for most of the total cross-section.
Hence the top-quark spin is strongly correlated with 
the positron spin direction determined in the top-quark rest-frame.

The production threshold for top-quarks of mass
175~GeV is far above the $Z$-boson pole. 
In this region, the $Z$ width is negligible and we can take the factors 
$f_{IJ}$ as real.
With real $f_{IJ}$'s one can choose a 
basis in which the $t_{\up} \bar t_{\up}$
and $t_{\down} \bar t_{\down}$
components vanish identically, see eqn.~(\ref{LRxsec}).
In this basis, which we refer to as the 
``\opt\ basis'', the spin angle $\sangle$ is given by
\beq
 \tan \sangle = { (f_{LL}+f_{LR}) ~\sqrt{1-\beta^2} ~\sin \thetas 
\over
f_{LL}~(\cos \thetas + \beta)+f_{LR}~(\cos \thetas - \beta) } \ 
\label{defnopt}
\eeq
and the polarized cross-sections 
are
\beqa
\ds ~(e^-_L~e^+_R \rightarrow t_{\up}~\tbar_{\up})  & = &
\ds ~(e^-_L~e^+_R \rightarrow t_{\down}~\tbar_{\down}) 
 = 0 \ , \nn \\[0.15in]
\ds ~(e^-_L~e^+_R \rightarrow t_{\up}~\tbar_{\down} 
 ~or~  t_{\down}~\tbar_{\up}) & = & 
\norme \Biggl[ ~f_{LL}(1 + \beta \cos \thetas) 
+ f_{LR}(1 - \beta \cos \thetas)
\nn \\[0.15in]
\lefteqn{ \pm  \sqrt{ 
(f_{LL}(1 + \beta \cos \thetas)
 +  f_{LR}(1 - \beta \cos \thetas))^2 
-4f_{LL}f_{LR}\beta^2 \sin^2 \thetas~}~ \Biggr] ^2 \ .}  \hspace{1.5in}
\label{Optimal}
\eeqa
For $|f_{LL}| \gg |f_{LR}|$ only the $t_{\up}\bar t_{\down}$ component
is substantially different from zero, and to leading order in 
$f_{LR} \over  f_{LL}$, the $t_{\down}\bar t_{\up}$ 
component is given by the same expression as for the Beamline 
basis\footnote{The polarized cross-section 
formulae~(\ref{LRxsec})
are also valid for top-quark pair production
in $q{\bar q}$-scattering, with an appropriate change of couplings.  
The \opt\ basis for $q{\bar q}$-scattering is 
given by eqn.~(\ref{defnopt}) with 
$f_{LL} = f_{LR} = f_{RL} = f_{RR}$, 
for which $\tan\sangle = {\sqrt{1-\beta^2}} \tan\thetas$ with
\newline
{\centerline{
$
\ds ~(q^-_L~ \bar{q}^+_R \rightarrow t_{\up}~\tbar_{\up} 
 ~or~  t_{\down}~\tbar_{\down})  = 0 $ } and }
{\centerline{
$
\ds ~(q^-_L~ \bar{q}^+_R \rightarrow t_{\up}~\tbar_{\down} 
 ~or~  t_{\down}~\tbar_{\up})  =  
\norms 
\Bigl[ 1 \pm \sqrt{1-\beta^2\sin^2 \thetas} ~\Bigr] ^2  \ .
$
}}
%
}.
In general there are two ``\opt'' bases for fermion pair production,
one for $e^-_L e^+_R$ scattering and the other for $e^-_R e^+_L$, since in
general $\frac{f_{LR}}{f_{LL}} \ne \frac{f_{RL}}{f_{RR}}$.
However for top-quark production 
the two ratios are approximately equal in both sign and magnitude, 
so that the two ``\opt''\ bases  are almost identical.
In the rest of this paper we will only use the \opt\ basis for $e^-_L e^+_R$
defined by eqn.~(\ref{defnopt}) even when discussing $e^-_R e^+_L$ scattering.

\begin{figure}[h]

\vspace*{9cm}
\includegraphics{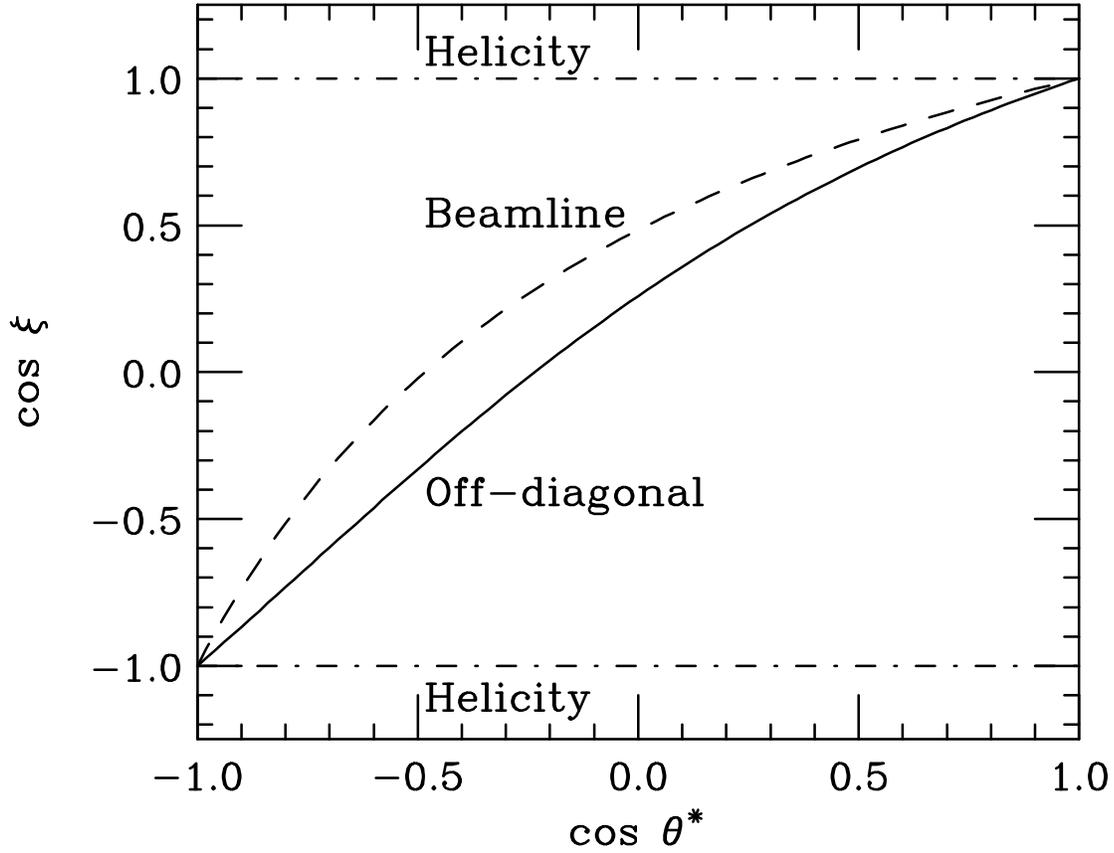}
\vspace{7.5cm}

\caption[]{
The dependence of the spin angle, $\sangle$, on the scattering angle, 
$\thetas$, for the helicity, Beamline and \opt\ 
(defined for $e^-_L~e^+_R$ scattering)
bases for a 175~GeV top-quark 
produced by a 400~GeV \epm\ collider.
}
\label{bases}
\end{figure}

To illustrate the different spin bases we now consider 
top-quark pair production, at a 400~GeV 
 \epm\ collider.
We take the top-quark mass to be 175~GeV ($\beta \sim 0.5$).
Fig.~\ref{bases} shows 
the dependence of the spin direction angle, 
$\sangle$, on the scattering angle, 
$\thetas$, for the helicity, Beamline and \opt\ bases.
The Beamline basis lies close to the 
\opt\ basis for all scattering angles.
For $\cos \thetas$ near zero there is a marked difference between
the helicity basis and the other two bases.
Note that as $\beta \rightarrow 1$, 
both the Beamline and 
\opt\ bases approach the helicity basis.
Therefore, for lepton colliders with center-of-mass energies greater 
than 1~TeV, all three bases are equivalent.

\begin{figure}[h]

\vspace*{10cm}
\includegraphics{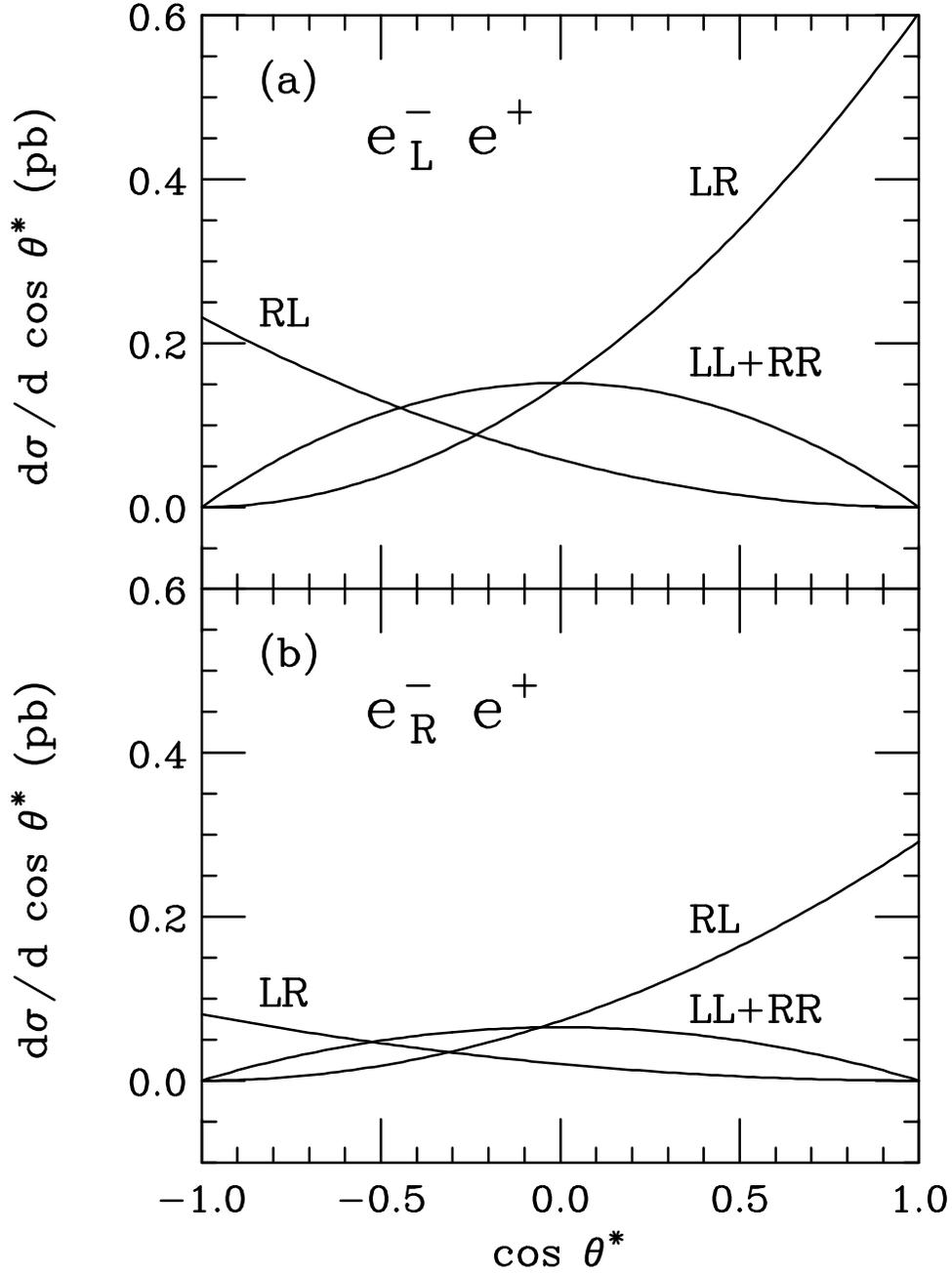}

\vspace{8.0cm}

\caption[]{
The differential cross-sections for producing top-quark pairs at a 400~GeV 
\epm\ collider
in the following  helicity configurations: $t_L {\bar t}_R$ (LR), 
$t_R {\bar t}_L$ (RL), and the sum of $t_L {\bar t}_L$ and 
$t_R {\bar t}_R$ (LL+RR), for left-handed and right-handed electron beams.
}
\label{helicitysig}
\end{figure}

\begin{figure}[h]

\vspace*{10cm}
\includegraphics{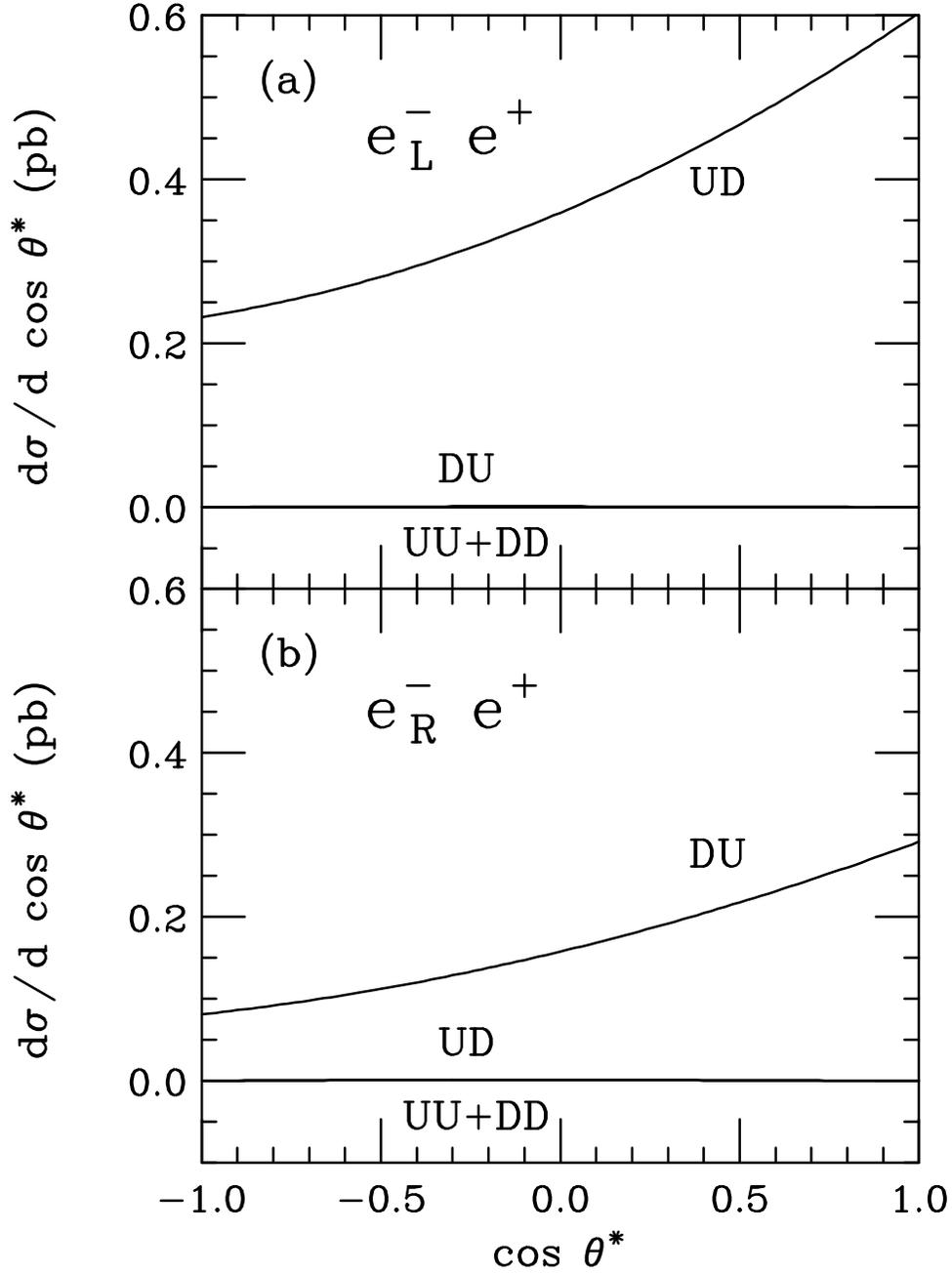}

\vspace{8.0cm}

\caption[]{
The differential cross-sections for producing top-quark pairs at a 400~GeV 
\epm\ collider
in the following spin configurations in the \opt\ basis 
(defined for $e^-_L~e^+_R$ scattering): 
$t_{\up} {\bar t}_{\down}$ (UD), 
$t_{\down} {\bar t}_{\up}$ (DU), and the sum of 
$t_{\up} {\bar t}_{\up}$  
and $t_{\down} {\bar t}_{\down}$
(UU+DD), for left-handed and right-handed electron beams.
}
\label{optimalsig}
\end{figure}

\begin{figure}[h]

\vspace*{10cm}
\includegraphics{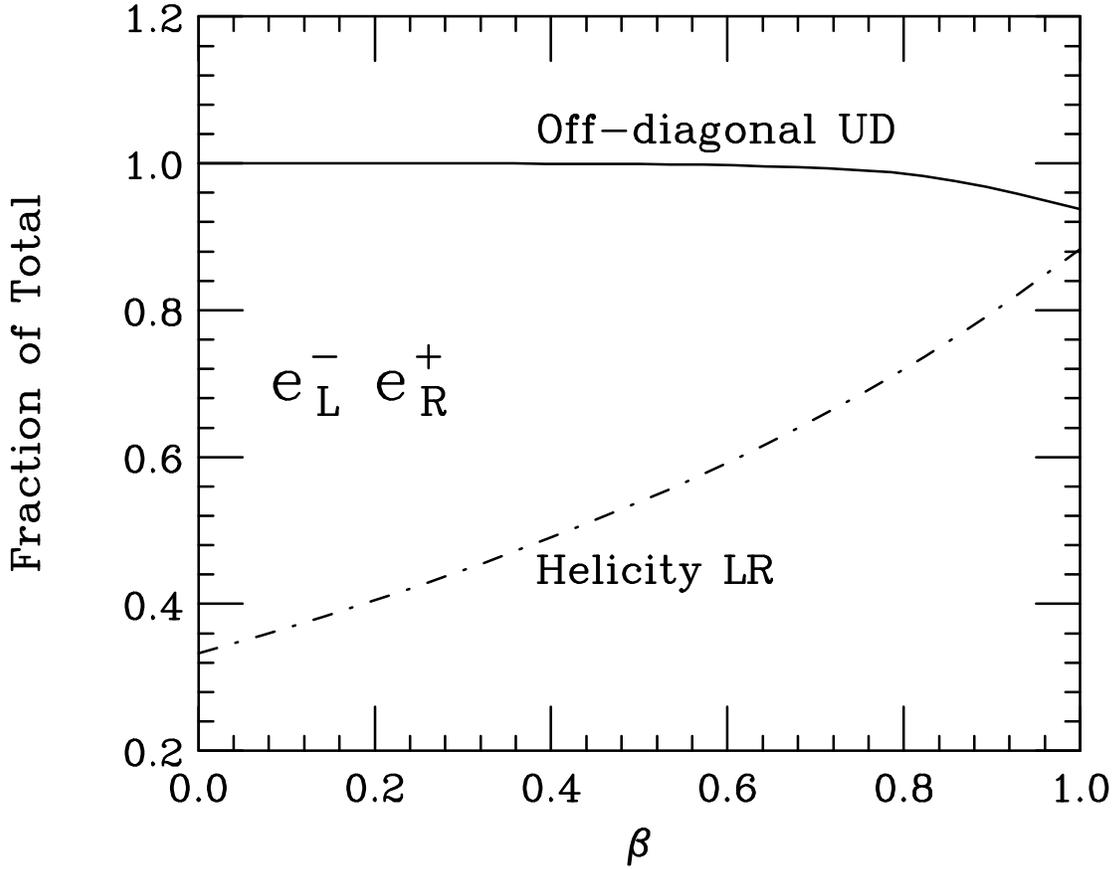}

\vspace{4.5cm}

\caption[]{
The fraction of top-quark pairs in the dominant spin configuration 
in the \opt\ basis and in the helicity basis, as a function of the top-quark
speed, $\beta$, in $e^-_L e^+_R$ scattering.
The solid line gives 
$\sigma(e^-_L e^+_R \rightarrow t_{\up} {\bar t}_{\down})/ 
\sigma_{T}$, defined in the \opt\ basis.
The dot-dashed line gives
 $\sigma(e^-_L e^+_R \rightarrow t_{L} {\bar t}_{R})/ 
\sigma_{T}$ in the helicity basis.
Here $\sigma_T$ is the total cross-section for
$e^-_L e^+_R \rightarrow t {\bar t}$. 
}
\label{beta}
\end{figure}

The cross-sections for producing $t {\bar t}$ pairs of definite helicities 
are shown in Fig.~\ref{helicitysig} for both \eL\ and \eR\ scattering.
While the dominant components of the signal are 
$t_L {\bar t}_R$ for a left-handed electron 
and  $t_R {\bar t}_L$ for a right-handed electron, 
other spin components make up more than 40\% of the total cross-section.
In the \opt\ basis, in contrast, only one spin 
component is appreciably non-zero for all values of the scattering angle;
\UD\ for \eL\ and \DU for \eR\ , see Fig.~\ref{optimalsig}.
All other components are more than two and a half 
orders of magnitude smaller than
these dominant contributions.
Thus, when defined in the \opt\ basis, the spins of the top quark and 
anti-quark 
produced in \eL\ and \eR\ scattering are essentially determined.
This continues to hold for top-quark pair production at higher energies.
In Fig.~\ref{beta} we show, for \eL\ collisions, 
the fraction of top-quark events in the
$t_{\up} {\bar t}_{\down}$ configuration, defined in the \opt\ basis,
as a function of the top-quark speed $\beta$. Also shown is the fraction 
of top-quark 
pairs in the $t_{L} {\bar t}_{R}$ helicity configuration. 
The \opt\ basis gives a very clean $t_{\up} {\bar t}_{\down}$ spin state
for all values of $\beta$ in these collisions. 
Similarly, the spin state $t_{\down} {\bar t}_{\up}$ dominates
\eR\ collisions at all energies.

We do not show here cross-section plots for the Beamline basis, 
since they are almost identical to those of Fig.~\ref{optimalsig}, 
except that the non-dominant contributions are now at the 1\% level
for a 400 GeV collider.
For both the Beamline basis and the \opt\ basis,
the contribution of higher order corrections to the 
non-dominant components is expected to increase 
their total contribution to the few percent level.


Since the top-quark pairs are produced in an unique spin configuration,
and the electroweak decay products of polarized top-quarks are 
strongly correlated to the spin axis, 
the top-quark events at  \epm\ collider have a
very distinctive topology. 
Deviations from this topology would signal anomalous couplings.
In the Standard Model, the predominant decay mode of the top-quark is 
$t \rightarrow b W^+$, with the $W^+$ 
decaying either hadronically or leptonically. 
For definiteness we consider here the decay 
$t \rightarrow b W^+\rightarrow b e^+ \nu$. 
The differential decay width of a polarized top-quark 
 depends non-trivially on three angles. 
The first is the angle, $\chiw$, between the top-quark spin and 
the  direction of motion of the $W$-boson in the top-quark rest-frame.
Next is the angle between the direction of 
motion of the $b$-quark and the positron in the W-boson rest-frame. 
We call this angle $\pi - \chie$. 
Finally, in the top-quark rest-frame, we have the azimuthal angle, $\pew$,
between the positron direction of motion and the top-quark spin 
around the direction of motion of the $W$-boson.

\begin{figure}[h]

\vspace*{9cm}
\includegraphics{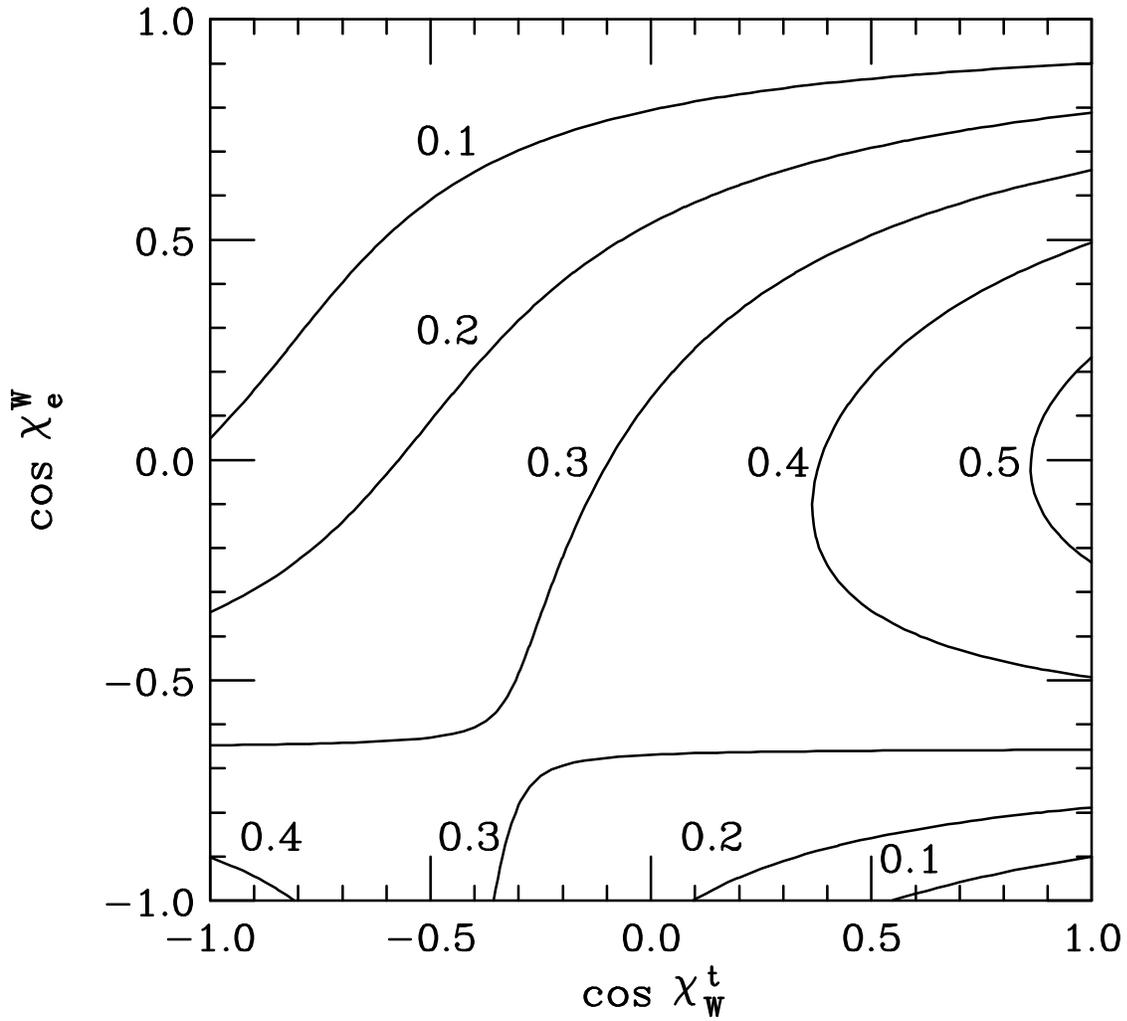}
\vspace{7.5cm}

\caption[]{
Contours of the top-quark decay distribution, eqn.~(\ref{diffgamma}), 
integrated over all $\pew$, in the  $\chie-\chiw$ plane. 
}
\label{gamma}
\end{figure}

\begin{figure}[h]

\vspace*{9cm}
\includegraphics{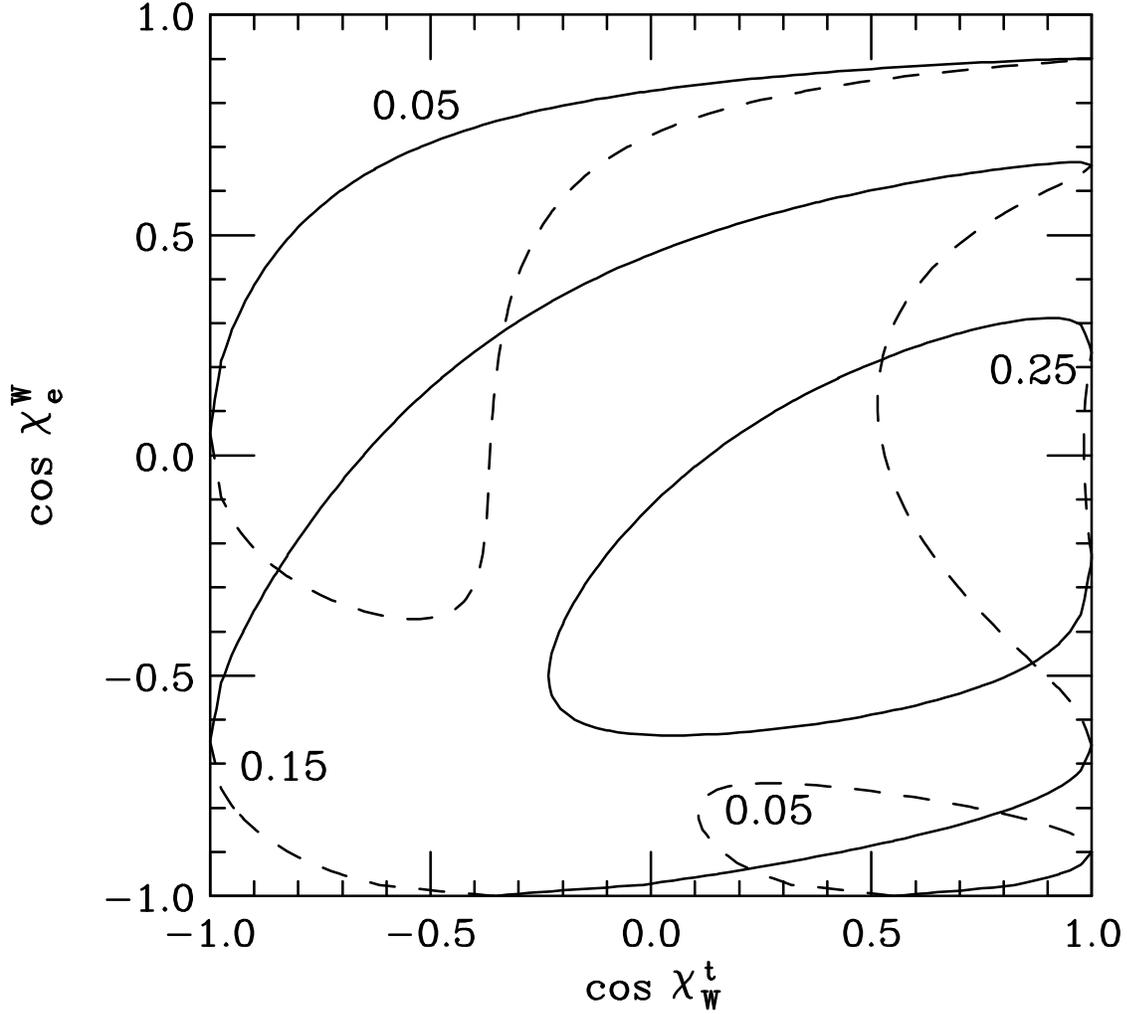}
\vspace{7.5cm}

\caption[]{
Contours of the top-quark decay distribution, eqn.~(\ref{diffgamma}), 
integrated over $\pew$ for $\cos\pew > 0$ (solid), 
and for $\cos\pew < 0$ (dashed) 
in the  $\chie-\chiw$ plane. 
The solid and dashed curves join continuously at the edges of the plot, 
where the interference term is zero.
}
\label{gammaphi}
\end{figure}

The differential polarized top-quark decay distribution in terms of 
these three angles is given by
\beqa
\frac{1}{\Gamma_T} ~{d^3 ~\Gamma \over d \cosw ~d \ce ~d \pew} & = &
\frac{3}{16 \pi ~(m_t^2+2m_W^2)} 
~\Bigl[m_t^2(1+\cosw)\sse \nn \\[0.2in]
+  m_W^2(1-\cosw)(1-\ce)^2 
& + & 2 m_t m_W (1-\ce)\se \sinw \cos \pew \Bigr] \ ,
\label{diffgamma}
\eeqa
where $m_t$ is the top-quark mass, $m_W$ is the $W$ mass,
and $\Gamma_T$ is the total decay width
(we neglect the  $b$-quark mass).
The first and second terms in~(\ref{diffgamma}) give the contributions 
of longitudinal and transverse $W$-bosons respectively. 
The interference term, given by the third 
term in~(\ref{diffgamma}), does not contribute to the total width, 
but its effects on the angular distribution of 
the top-quark decay products are sizable.
Fig.~\ref{gamma} shows 
contour plots 
of the differential angular decay distribution
in the $\chie-\chiw$ plane\footnote{We take $M_W=80$~GeV.},
after integrating over the azimuthal angle 
$\pew$.
Fig.~\ref{gammaphi} shows analogous contours integrated over $\pew$ 
for positive (solid lines) and for negative (dashed lines) 
values of $\cos\pew$ separately. 
The pronounced difference between these is related to the 
size of the interference term, 
which can be seen from the $\pew$-distribution 
\beqa
\frac{1}{\Gamma_T} ~{d ~\Gamma \over d \pew} & = &
\frac{1}{2 \pi }
~\Bigl[1 + \frac{3\pi^2 m_t m_W}{16(m_t^2+2m_W^2)} \cos \pew \Bigr] \ .
\label{pewgamma}
\eeqa
For a 175 GeV top-quark the coefficient in front of the 
cosine term has a value equal to 0.59, 
therefore the maximum and minimum values of this distribution are approximately
4 to 1.

There are also significant correlations 
of the angle between the top-quark spin  
and the momentum of the $i$-th decay product, $\chi^t_i$,
measured in the top-quark rest-frame.
The differential decay rate of the top-quark is given by
\beqa
\frac{1}{\Gamma_T} ~{d ~\Gamma \over d \cos \chi^t_i} & = &
\frac{1}{2}
~\Bigl[1 + \alpha_i \cos \chi^t_i \Bigr] \ ,
\eeqa
where $\alpha_b = -0.41$, $\alpha_{\nu} = -0.31$ and $\alpha_{e^+} = 1$,
for $m_t = 175$ GeV,
see ref.~\cite{Jezabek2}.


In summary, we have presented simple analytic expressions for the polarized 
cross-section for top-quark pair production in polarized  \epm\ colliders.
For a particular choice of axes, the \opt\ basis, 
not only do the like-spin contributions vanish,
but one spin configuration dominates the total cross-section.
In this configuration, the top-quark spin is strongly correlated with 
the positron spin direction determined in the top-quark rest-frame.
The subsequent electroweak decays of the top-quark pair give decay 
products whose angular distributions are highly correlated with the parent
top-quark spin. 
Top-quark pair events thus have a distinctive topology.
This topology is sensitive to  the  top-quark couplings to the $Z$-boson
and to the photon, which determine    
the orientation and the size of the top-quark and top anti-quark 
polarizations,  
as well as to the top-quark couplings to the $W$ and the $b$-quark, 
which determine its decay distributions. 
Angular correlations in top-quark events 
 may therefore be used to constrain deviations 
from the Standard Model.
We have also shown that the interference between the {\it longitudinal}
and {\it transverse}
$W$-bosons has a significant impact on the angular distribution 
of the top-quark decay products, and thus will 
provide additional means for 
testing the Standard Model predictions for top-quark decays.


\acknowledgements

The Fermi National Accelerator 
Laboratory is operated by Universities Research Association,
Inc., under contract DE-AC02-76CHO3000 with the U.S. Department
of Energy.



\newpage

\newpage


\newpage


\vspace*{0.5cm}
 

\end{document}